\documentclass[pra,twocolumn,showpacs,preprintnumbers,superscriptaddress]{revtex4}
\usepackage{xcolor}
\usepackage{times}
\usepackage{bm}
\usepackage{float}
\usepackage{graphicx}
\usepackage{amsbsy}
\usepackage{amsmath}
\usepackage{amsfonts}
\usepackage{amsthm}

\begin{document}

\theoremstyle{plain}
\newtheorem{theorem}{Theorem}
\newtheorem{lemma}[theorem]{Lemma}
\newtheorem{corollary}[theorem]{Corollary}
\newtheorem{proposition}[theorem]{Proposition}
\newtheorem{conjecture}[theorem]{Conjecture}

\theoremstyle{definition}
\newtheorem{definition}[theorem]{Definition}

\theoremstyle{remark}
\newtheorem*{remark}{Remark}
\newtheorem{example}{Example}
\title{Teleportation criteria based on maximum eigenvalue of the shared $d\otimes d$ dimensional mixed state: Beyond Singlet Fraction}
\author{Anuma Garg, Satyabrata Adhikari}
\email{anumagarg_phd2k18@dtu.ac.in, satyabrata@dtu.ac.in} \affiliation{Delhi Technological
University, Delhi-110042, Delhi, India}

\begin{abstract}
\centerline{Abstract}
We derive a criteria for the detection of $d\otimes d$ dimensional negative partial transpose (NPT) entangled state useful for teleportation. The newly derived criteria are based on the maximum eigenvalue of the NPT entangled state, which is easier to determine experimentally than to completely reconstruct the state via tomography. We then illustrate our criteria by considering a class of qubit-qubit system and qutrit-qutrit system.
\end{abstract}
 \maketitle

\section{Introduction}
Quantum teleportation is an important topic to study in quantum information science. It plays a vital role in the development of
quantum information theory and quantum technologies \cite{nielsen,wilde}. Bennett et.al. \cite{Bennett1} have developed the first protocol of quantum teleportation for two-qubit system. The developed protocol talks about the transfer of information contained in a qubit from a sender (say, Alice) to a receiver (say, Bob). To execute this protocol, firstly Alice and Bob shared an entangled state between them. Then she perform two-qubit Bell-state measurement on particles in her possession. After that, she communicates the measurement result to Bob by sending two classical bits. The receiver Bob then reconstructs the quantum state at his place by applying suitable unitary operation such as $I, \sigma_{x}, \sigma_{y}, \sigma_{z}$ on his qubit according to the measurement outcome sent by the sender Alice.\\
 The first protocol of quantum teleportation can be considered as a basic protocol for other quantum schemes such as quantum repeater \cite{briegel}, quantum gate teleportation \cite{gottesman}, port-based teleportation \cite{ishizaka1}. We should note that the most important ingredient in quantum teleportation protocol is the resource state, which is shared between two distant parties because it would not be possible to realize quantum teleportation without shared entangled state. In a realistic situation, it is not possible to keep the shared entangled state in a pure form as the particle has to interact with the environment. Thus, the shared entangled state in general become a mixed entangled state. In this context, the question may arise that whether the generated mixed entangled state is useful as a resource state in quantum teleportation? To answer this question, a quantity known as singlet fraction has been defined \cite{horodecki1}. It is defined as the maximum overlap between the quantum state described by the density operator $\rho$ and a maximally entangled state in a finite dimensional Hilbert space. The usefulness of the shared entangled state between two distant partners in a teleportation protocol depends on the value of the singlet fraction. For $d\otimes d$ dimensional system, the shared state is useful in quantum teleportation if the singlet fraction of the shared state is greater than $\frac{1}{d}$.\\
 The singlet fraction also characterizes the nature of the quantum state in the sense that if the given state is separable then the singlet fraction of the given two-qubit mixed state is less than or equal to $\frac{1}{2}$ \cite{horodecki1}. In another way, it can be stated as if the singlet fraction of an arbitrary given state is greater than $\frac{1}{2}$ then the state is entangled. But the converse of the statement is not true. This means that there exist two-qubit mixed entangled state whose singlet fraction is less than or equal to $\frac{1}{2}$ and hence not useful in quantum teleportation. In this perspective, Badziag et.al. \cite{badziag} have shown that a dissipative interaction with the environment is sufficient to improve the value of the singlet fraction. They have presented a class of entangled quantum state whose singlet fraction is exactly equal to $\frac{1}{2}$ before interaction with the environment but after the interaction with the environment, the value of singlet fraction improves. Even getting this result also, the question remains that whether interaction with the environment increases the singlet fraction of any two-qubit mixed state? The answer is in affirmative. Verstraete et al. \cite{verstraete} have studied this problem and obtained trace preserving LOCC that enhances the singlet fraction and make its value greater than $\frac{1}{2}$ for any two-qubit mixed entangled state. They have derived a connection between the optimal singlet fraction and partial transpose of a given state. The established relation tells us that the two-qubit state is useful as a resource state for teleportation if and only if the optimal singlet fraction is greater than $\frac{1}{2}$.\\
Till now, we have discussed about the resource state useful in teleportation for $2\otimes 2$ dimensional system. We now continue our discussion with higher dimensional system. Generally, it has been proved that to teleport an arbitrary d-dimensional pure state, only maximally entangled pure state in $d\otimes d$ is required \cite{albeverio}. Zhao et.al \cite{zhao} have derived the necessary and sufficient conditions of faithful teleportation of an arbitrary d-dimensional pure state with $m\times d$ and $d\times n$ dimensional entangled resource, where $m$ and $n$ denoting the dimension of the first qubit in $m\times d$ and second qubit in $d\times n$ dimensional entangled resource states respectively. A general expression for the output state of the quantum channel associated with the original teleportation protocol with an arbitrary $d\otimes d$ dimensional mixed resource state has been
obtained in \cite{bowen}.\\
The motivation of this work is two fold: (i) Since partial transposition is a non-physical operation and cannot be implemented in a laboratory so
we apply SPA on partial transposition of the given state. Being SPA a completely positive map, the expression of singlet fraction get free from partial transposition operation and thus it can be implemented in an experiment. (ii) The second motivation comes from the problem of estimation of singlet fraction in higher dimensional bipartite system. Since singlet fraction depends on bipartite maximally entangled states but in higher dimensional system, it is very difficult to construct bipartite maximally entangled state so it would not be an easy task to get the experimentally estimated value of singlet fraction for higher dimensional bipartite system. Thus, it is desirable to establish another criteria for the detection of entangled state useful in teleportation and that must be easy to implement in experiment. Therefore, instead of singlet fraction if the criterion is expressed in terms of the eigenvalue then it requires lesser number of measurements than to completely reconstruct the state via tomography \cite{ekert}. Thus, in terms of number of required measurements, the criterion based on eigenvalue is more efficient than quantum tomography.\\
This paper is organized as follows: In section-II, we state some preliminary result that will be needed in the later sections. In section-III,
we revisit Verstraete et.al. work and modify their result in terms of SPA-PT of a given quantum state. In section-IV, we give the criterion in terms of maximum eigenvalue for the detection of $d\otimes d$ dimensional NPT entangled resource state useful in quantum teleportation. In section-V, we will show that the criterion based on eigenvalue may fail and thus we provide another criterion based on Dembo's bound. Lastly, we conclude in section-VI.


\section{Preliminary Results}
\textbf{Result 1:} For any two Hermitian $d\times d$ matrices A and B, we have \cite{Lasserre,anu}
\begin{eqnarray}
\lambda_{min}(A)Tr(B)\leq Tr(AB)\leq \lambda_{max}(A)Tr(B)
\label{result1}
\end{eqnarray}
where the eigenvalues of A are arranged as $\lambda_{min}=\lambda_{1}\leq \lambda_{2}\leq \lambda_{3}\leq.......\leq \lambda_{d}=\lambda_{max}$.\\
\textbf{Result 2:} An arbitrary $d\otimes d$ dimensional entangled quantum state $\rho$ is useful for quantum teleportation if \cite{badziag}
\begin{eqnarray}
F^{max}(\rho)>\frac{1}{d}
\label{result2}
\end{eqnarray}
where $F^{max}(\rho)$ denote the singlet fraction of $\rho$ and it is given by \cite{liang}
\begin{eqnarray}
F^{max}(\rho) &=& max_{U_{A},U_{B}} \{ F[\rho, (U_{A}\otimes U_{B})|\phi_{d}^{+}\rangle\langle\phi_{d}^{+}| \nonumber\\&&
(U_{A}^{\dagger}\otimes U_{B}^{\dagger})]\}
\label{defsingfrac}
\end{eqnarray}
where $|\phi_{d}^{+}\rangle=\frac{1}{\sqrt{d}}\sum_{i=0}^{d-1}|ii\rangle$.\\
It is known that there exist an entangled state $\rho$ for which $F^{max}(\rho)\leq \frac{1}{d}$ but these entangled states may or may not be useful for teleportation.\\
\textbf{Result 3:} In standard teleportation scheme, the relation between the singlet fraction of $\rho$ in $d\otimes d$ dimensional Hilbert space and maximal achievable teleportation fidelity $f^{tel}(\rho)$ is given by \cite{horodecki1}
\begin{eqnarray}
f^{tel}(\rho)=\frac{dF^{max}(\rho)+1}{d+1}
\label{reltelfidsingfrac}
\end{eqnarray}
The teleportation fidelity can be regarded as useful if $f^{tel}(\rho)>\frac{2}{d+1}$, where the expression $\frac{2}{d+1}$ denoting
the classical teleportation fidelity.\\
\textbf{Result 4:} Let us consider an arbitrary two qubit state described by the density operator $\rho_{12}$
\begin{eqnarray}
\rho_{12}=
\begin{pmatrix}
  e_{11} & e_{12} & e_{13} & e_{14} \\
  e_{12}^{*} & e_{22} & e_{23} & e_{24} \\
  e_{13}^{*} & e_{23}^{*} & e_{33} & e_{34} \\
  e_{14}^{*} & e_{24}^{*} & e_{34}^{*} & e_{44}
\end{pmatrix}, \sum_{i=1}^{4}e_{ii}=1
\end{eqnarray}
where $(*)$ denotes the complex conjugate, then the structural physical approximation (SPA) of partial transpose (PT) of $\rho_{12}$ is given by \cite{adhikari1}
\begin{eqnarray}
\tilde{\rho}_{12}&=&[\frac{1}{3}(I\otimes\widetilde{T})+\frac{2}{3}(\widetilde{\Theta}\otimes D)](\rho_{12})\nonumber\\&=&
\begin{pmatrix}
  E_{11} & E_{12} & E_{13} & E_{14} \\
  E_{12}^{*} & E_{22} & E_{23} & E_{24} \\
  E_{13}^{*} & E_{23}^{*} & E_{33} & E_{34} \\
  E_{14}^{*} & E_{24}^{*} & E_{34}^{*} & E_{44}
\end{pmatrix}
\label{spa1}
\end{eqnarray}
where
\begin{eqnarray}
&&E_{11}=\frac{1}{9}(2+e_{11}),E_{12}=\frac{1}{9}e_{12}^{*}, E_{13}=\frac{1}{9}e_{13},\nonumber\\&&
E_{14}=\frac{1}{9}e_{23}, E_{22}=\frac{1}{9}(2+e_{22}),E_{23}=\frac{1}{9}e_{14},\nonumber\\&&
E_{24}=\frac{1}{9}e_{24},E_{33}=\frac{1}{9}(2+e_{33}),E_{34}=\frac{1}{9}e_{34}^{*},\nonumber\\&&
E_{44}=\frac{1}{9}(2+e_{44})
\label{spa2a}
\end{eqnarray}
$\widetilde{T}$ is the SPA of transpose map $T$ and $\widetilde{\Theta}$ denotes the SPA of inversion map $\Theta$. The inversion map $\Theta$ is defined as $\Theta(\rho)$=-$\rho$ and $D(\rho)=\frac{I_{2}}{2}$ denotes
the polarization. SPA-PT for two qubit photonic system using single-photon polarization qubits and linear optical devices has been demonstrated in \cite{lim}.\\
\textbf{Result 5:} For any $n\otimes n$ Hermitian positive semi-definite operator $R_{n}$ with eigenvalues $\lambda_{1}\leq \lambda_{2}\leq.....\lambda_{n}$, Dembo's bound \cite{dembo,park} is given by
\begin{eqnarray}
&&\frac{c+\eta_{1}}{2}+\sqrt{\frac{(c-\eta_{1})^{2}}{4}+(b^{*})^{T}b}\leq \lambda_{n}(R_{n}) \nonumber\\&\leq& \frac{c+\eta_{n-1}}{2}+\sqrt{\frac{(c-\eta_{n-1})^{2}}{2}+(b^{*})^{T}b}
\label{db1}
\end{eqnarray}
where  $R_{n}=\begin{pmatrix}
  R_{n-1} & b \\
  (b^{*})^{T} & c
  \end{pmatrix}$,
$\eta_{1}$ is the lower bound on minimal eigenvalue of $R_{n-1}$, $\eta_{n-1}$ is the upper bound on maximal eigenvalue of $R_{n-1}$
and $b$ is a vector of dimension $n-1$.
\section{Revisiting maximal singlet fraction of mixed two-qubit state}
\noindent Verstraete et.al. \cite{verstraete} have derived the optimal trace-preserving local operation together with classical communication and have shown that it optimally increases the singlet fraction of a mixed quantum state $\rho_{12}$ and hence maximize its teleportation fidelity. They have studied the case of two-qubit system and proved that if the state is entangled then it is always possible to increase the singlet fraction above $\frac{1}{2}$ and thus make teleportation fidelity greater than $\frac{2}{3}$. Since their result is based on the partial transposition operation so it is not possible to realize it in the experimental setup. In this section, we revisit their result and apply structural physical approximation (SPA) of partial transposition. By doing this, the final expression of the singlet fraction get modified and since SPA of partial transposition is a completely positive map so the singlet fraction can be estimated experimentally. However, we find that in this case the value of the singlet fraction is not always greater than $\frac{1}{2}$. Therefore, the result of this section motivate us further to investigate for new teleportation criteria that will be studied in the following section.\\
The expression of optimal singlet fraction after LOCC for a two-qubit state $\rho_{12}$ is given by \cite{verstraete}
\begin{eqnarray}
F^{opt}_{LOCC}(\rho_{12})=\frac{1}{2}-Tr(X^{opt}\rho_{12}^{\Gamma})
\label{optsingfrac}
\end{eqnarray}
where without any loss of generality, we denote $\Gamma$ as the partial transposition with respect to the second subsystem, and $X^{opt}$ is described by a $4\times 4$ matrix, which will be of rank one and it is expressed in the form as
\begin{eqnarray}
X^{opt}=(A\otimes I_{2})|\psi\rangle_{12}\langle\psi|(A^{\dagger}\otimes I_{2})
\label{optimalfilter}
\end{eqnarray}
with $|\psi\rangle_{12}=\frac{1}{\sqrt{2}}(|00\rangle+|11\rangle)$, $A$ denotes a filter described by a $2\times 2$ matrix and $I_{2}$ represents a $2\times 2$ identity matrix.\\
It is clear from equation (\ref{optsingfrac}) that if $\rho_{12}$ is an entangled state then
$F^{opt}_{LOCC}(\rho_{12})$ always greater than $\frac{1}{2}$ and hence one can reach the conclusion that
every mixed two-qubit entangled state is useful in teleportation. But an important point to note here is that the partial transposition $\Gamma$ is not a completely positive map so $\rho_{12}^{\Gamma}$ is not a physically realizable operation. Thus the value of $F^{opt}_{LOCC}(\rho_{12})$ cannot be realized in experiment. To overcome this problem, we use SPA-PT of $\rho_{12}$ and re-express $Tr(X^{opt}\rho_{12}^{\Gamma})$
given in (\ref{optsingfrac}) as \cite{adhikari1}
\begin{eqnarray}
Tr(X^{opt}\rho_{12}^{\Gamma})=9Tr(X^{opt}\widetilde{\rho}_{12})-2
\label{spa1}
\end{eqnarray}
where $\widetilde{\rho}_{12}$ is the SPA-PT of $\rho_{12}$.\\
Using (\ref{spa1}), equation (\ref{optsingfrac}) can be re-expressed as
\begin{eqnarray}
F^{opt}_{LOCC}(\rho_{12})=\frac{5}{2}-9Tr(X^{opt}\widetilde{\rho}_{12})
\label{optsingfrac1}
\end{eqnarray}
Now it is possible to estimate the value of $F^{opt}_{LOCC}(\rho_{12})$ experimentally but we have to pay cost in a way that the obtained value of $F^{opt}_{LOCC}(\rho_{12})$ may not be always greater than $\frac{1}{2}$. If it may happen that $Tr(X^{opt}\widetilde{\rho}_{12})<\frac{2}{9}$ then only $F^{opt}_{LOCC}(\rho_{12})$ is greater than $\frac{1}{2}$. This implies that if there exist state for which $Tr(X^{opt}\widetilde{\rho}_{12})\geq\frac{2}{9}$  then $F^{opt}_{LOCC}(\rho_{12})\leq\frac{1}{2}$. To illustrate, let us consider a two-qubit state described by the density operator
\begin{eqnarray}
\sigma_{12}=
\begin{pmatrix}
0 & 0 & 0 & 0\\
0 & b & f & 0 \\
0 & f^{*} & d & 0 \\
0 & 0 & 0 & e
\end{pmatrix},~~~b+d+e=1
\label{state1}
\end{eqnarray}
where $*$ denote the complex conjugation.\\
The density matrix $\sigma_{12}$ has been studied by many authors in different contexts \cite{bruss}-\cite{ishizaka}. The state $\sigma_{12}$ is an entangled state and its concurrence is given by \cite{wootters,bruss}
\begin{eqnarray}
C(\sigma_{12})=2|f|
\label{concurrence}
\end{eqnarray}
Using Result 4, we can obtain the SPA-PT of $\sigma_{12}$ as
\begin{eqnarray}
\widetilde{\sigma}_{12}=
\begin{pmatrix}
\frac{2}{9} & 0 & 0 & \frac{f}{9}\\
0 & \frac{2+b}{9} & 0 & 0 \\
0 & 0 & \frac{2+d}{9} & 0 \\
\frac{f^{*}}{9} & 0 & 0 & \frac{2+e}{9}
\end{pmatrix}
\label{spasigma}
\end{eqnarray}
Now if we consider the filter $A$ of the form as $\begin{pmatrix}
a & 0 \\
0 & 1
\end{pmatrix}, 0\leq a\leq 1$, then $X^{opt}$ is given by
\begin{eqnarray}
X^{opt}=
\begin{pmatrix}
\frac{a^{2}}{2} & 0 & 0 & \frac{a}{2}\\
0 & 0 & 0 & 0 \\
0 & 0 & 0 & 0 \\
\frac{a}{2} & 0 & 0 & \frac{1}{2}
\end{pmatrix}
\label{optfilter1}
\end{eqnarray}
The optimal singlet fraction of $\sigma_{12}$ is given by
\begin{eqnarray}
F^{opt}_{LOCC}(\sigma_{12})=\frac{5}{2}-9Tr(X^{opt}\widetilde{\sigma}_{12})
\label{optsingfrac10}
\end{eqnarray}
where $Tr(X^{opt}\widetilde{\sigma}_{12})$ is given by
\begin{eqnarray}
Tr(X^{opt}\widetilde{\sigma}_{12})= \frac{2a^{2}+2aRe(f)+2+e}{18}
\label{opttracesigma}
\end{eqnarray}
The inequality $\frac{2a^{2}+2aRe(f)+2+e}{18} \geq \frac{2}{9}$ holds if the filtering parameter $a$ satisfies
\begin{eqnarray}
\frac{-Re(f)+\sqrt{Re(f)^{2}-2e+4}}{2}\leq a
\label{parameterrange2}
\end{eqnarray}
Therefore, $F^{opt}_{LOCC}(\sigma_{12})$ is less than equal to $\frac{1}{2}$ iff (\ref{parameterrange2}) holds.\\
Let us now consider a particular case where we can set the values of the state parameter as: $b=0.2$, $d=0.4$, $e=0.4$ and $f=0.25+0.1i$. Using these values, the density matrix given in (\ref{state1}) reduces to
\begin{eqnarray}
\sigma^{(1)}_{12}=
\begin{pmatrix}
0 & 0 & 0 & 0\\
0 & 0.2 & 0.25+0.1i & 0 \\
0 & 0.25-0.1i & 0.4 & 0 \\
0 & 0 & 0 & 0.4
\end{pmatrix}
\label{stateexample1}
\end{eqnarray}
The SPA-PT of $\sigma^{(1)}_{12}$ is given by
\begin{eqnarray}
\tilde{\sigma}^{(1)}_{12}=
\begin{pmatrix}
\frac{2}{9} & 0 & 0 & \frac{0.25+0.1i}{9}\\
0 & \frac{2.2}{9} & 0 & 0 \\
0 & 0 & \frac{2.4}{9} & 0 \\
\frac{0.25-0.1i}{9} & 0 & 0 & \frac{2.4}{9}
\end{pmatrix}
\label{spaptstateexample1}
\end{eqnarray}
The value of $Tr(X^{opt}\widetilde{\sigma}^{(1)}_{12})$ is given by
\begin{eqnarray}
Tr(X^{opt}\widetilde{\sigma}^{(1)}_{12})= \frac{2a^{2}+0.5a+2.4}{18}
\label{optsingfracsigmaexample1}
\end{eqnarray}
Thus, the optimal singlet fraction of $\sigma^{(1)}_{12}$ is given by
\begin{eqnarray}
F^{opt}_{LOCC}(\sigma^{(1)}_{12})=\frac{2.6-2a^{2}-0.5a}{2}, 0.78\leq a \leq 1
\label{optsingfrac10example1}
\end{eqnarray}
\begin{figure}[h!]
\centering
\includegraphics[scale=0.70]{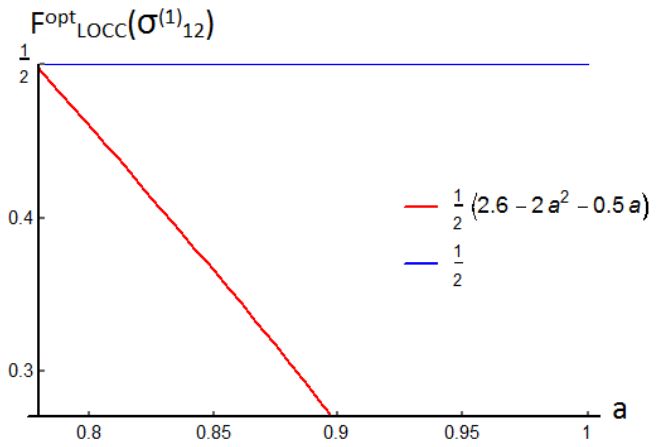}
\caption{Plot of optimal singlet fraction obtained after LOCC operation versus the filtering parameter a.}
\end{figure}
Figure 1 illustrate the fact that $F^{opt}_{LOCC}(\sigma^{(1)}_{12})$ is always less than or equal to $\frac{1}{2}$.\\
This motivates us to investigate for new teleportation criteria which may go beyond singlet fraction and identify not only two-qubit
entangled state but also higher dimensional NPT entangled state, which may be useful in quantum teleportation. We will study this in the later sections.
\section{Teleportation Criteria in terms of maximum eigenvalue}
\noindent In this section, we derive a criterion for the usefulness of the shared $d\otimes d$ dimensional NPT entangled states in quantum teleportation. In particular, for $2\otimes 2$ dimensional entangled states, the derived criterion may be useful in a situation when singlet fraction calculated before sending a qubit through the local environment is less than or equal to $\frac{1}{2}$. This means that if we don't use local environment to increase the value of singlet fraction, our criterion can still detect whether the shared resource state is useful for teleportation or not.\\
\textbf{Lemma 1:}
The maximum eigenvalue of an arbitrary $d\otimes d$ dimensional quantum state $\rho$ is always greater than or equal to the singlet fraction of
$\rho$. Mathematically, it can be expressed as
\begin{eqnarray}
\lambda_{max}(\rho)\geq F^{max}(\rho)
\label{maxeigsingfrac}
\end{eqnarray}
where $\lambda_{max}(\rho)$ denote the maximum eigenvalue of $\rho$.\\
\textbf{Proof:} Let us start with the definition of the singlet fraction given in (\ref{defsingfrac}), which can be
re-expressed as
\begin{eqnarray}
F^{max} (\rho)&=&max_{U_{A},U_{B}} Tr[\rho(U_{A} \otimes U_{B})|\phi_{d}^{+}\rangle\langle\phi_{d}^{+}| (U_{A}^{\dagger}\otimes U_{B}^{\dagger})]\nonumber\\
&=& max_{U_{A},U_{B}} Tr[(U_{A}^{\dagger}\otimes U_{B}^{\dagger})\rho(U_{A} \otimes U_{B})|\phi_{d}^{+}\rangle\langle\phi_{d}^{+}|]\nonumber\\
&\leq&  \{max_{U_{A},U_{B}} \lambda_{max}[(U_{A} \otimes U_{B})\rho(U_{A}^{\dagger}\otimes U_{B}^{\dagger})]\}\nonumber\\&& \{Tr[|\phi_{d}^{+}\rangle\langle\phi_{d}^{+}|]\} \nonumber\\ &=&  max_{U_{A},U_{B}} \lambda_{max}[(U_{A} \otimes U_{B})\rho(U_{A}^{\dagger}\otimes U_{B}^{\dagger})]\nonumber\\ &=&\lambda_{max}(\rho)
\label{criterion1}
\end{eqnarray}
Hence proved.\\
The inequality in the third step is a consequence of the result 1 and the last equality follows from a well known fact that the two quantum states $(U_{A} \otimes U_{B})\rho(U_{A}^{\dagger}\otimes U_{B}^{\dagger})$ and $\rho$ have same set of eigenvalues \cite{ziman}. Therefore, we have provided the alternative proof of this result that has already been obtained in \cite{juan}.\\
\textbf{Lemma 2:} If $\lambda_{max}(\rho)$ denotes the maximum eigenvalue of $d\otimes d$ dimensional quantum state $\rho$, then
\begin{eqnarray}
\frac{1}{d^{2}}\leq \lambda_{max}(\rho)\leq 1
\label{critmaxeig}
\end{eqnarray}
Let us consider the noisy singlet state of the form
\begin{eqnarray}
\rho_{p}=p|\phi^{+}\rangle\langle\phi^{+}|+(1-p)\frac{I\otimes I}{d^{2}}, 0\leq p\leq 1
\label{noisysinglet}
\end{eqnarray}
where $|\phi^{+}\rangle=\frac{1}{\sqrt{d}} \sum_{i=0}^{d-1}|ii\rangle$.\\
The maximum eigenvalue of the density matrix $\rho_{p}$ is given by
\begin{eqnarray}
\lambda_{max}(\rho_{p})&=&\lambda_{max}[p|\phi^{+}\rangle\langle\phi^{+}|+(1-p)\frac{I\otimes I}{d^{2}}], 0\leq p\leq 1
\nonumber\\&=& p+\frac{1-p}{d^{2}}
\label{eigennoisysinglet}
\end{eqnarray}
Now, since it is known that the state $\rho_{p}$ is separable if and only if $0\leq p\leq \frac{1}{d+1}$ \cite{horodecki2} so we can say that the state $\rho_{p}$ is separable if and only if $\lambda_{max}(\rho_{p})\leq \frac{1}{d}$.\\
Therefore, we can state the following theorem.\\
\textbf{Theorem:} An arbitrary $d\otimes d$ dimensional isotropic quantum state $\rho_{p}$ shared between two distant partners is separable if and only if
\begin{eqnarray}
\lambda_{max}(\rho_{p})\leq \frac{1}{d}
\label{critmaxeig}
\end{eqnarray}
It should be noted that the above condition is necessary and sufficient condition only for isotropic state.\\
\textbf{Corollary 1:} An entangled quantum state $\rho$ shared between two distant partners is useful for quantum teleportation if
\begin{eqnarray}
\lambda_{max}(\rho) > \frac{1}{d}
\label{critmaxeig}
\end{eqnarray}
\textbf{Corollary 2:} The upper bound of the maximum achievable teleportation fidelity from a given bipartite state $\rho$
in $d\otimes d$ dimensional Hilbert space is given by
\begin{eqnarray}
f^{tel}(\rho)\leq \frac{\lambda_{max}(\rho)d+1}{d+1}
\label{telfid}
\end{eqnarray}
\subsection{Examples}
\noindent We are now in a position to discuss a few examples in which our criterion detects mixed entangled states useful in quantum teleportation while singlet fraction failed to do so. In this, we will study a few quantum states $\rho$ for which singlet fraction $F_{max}(\rho)$ and maximum eigenvalue $\lambda_{max}(\rho)$ satisfies the inequality
\begin{eqnarray}
F_{max}(\rho)\leq \frac{1}{d}< \lambda_{max}(\rho)
\label{ineq1}
\end{eqnarray}
When The left part of the inequality (\ref{ineq1}) holds, then We are uncertain about the usefulness of state $\rho$ for a resource state in quantum teleportation except for d=2, where we can affirmatively say that state $\rho$ is not useful for a resource state in quantum teleportation. If the right part of the inequality (\ref{ineq1}) holds true, then we can say that the state $\rho$ can be used in teleportation. This means that when singlet fraction is unable to detect the useful states for quantum teleportation, then the maximum eigenvalue can serve the purpose. Also we should note that for the above quantum state $\rho$, interaction with the environment is not taken into account.\\
\subsubsection{Example 1}
Let us now recall the quantum state described by the density operator $\sigma_{12}^{(1)}$ given in
(\ref{stateexample1}). Firstly, we need to check whether the state is entangled. Let us consider a witness operator of the form\cite{Riccardi}
\begin{eqnarray}
W_{1}&=& \frac{1}{4}( I \otimes I + I\otimes \sigma_{z} +\sigma_{z} \otimes I-\sigma_{x} \otimes \sigma_{x}+\nonumber\\&& \sigma_{y}\otimes \sigma_{y})
\end{eqnarray}
where $\sigma_{x}, \sigma_{y}, \sigma_{z}$ are denoting pauli matrices.\\
The expectation value of the operator $W_{1}$ with respect to the state $\sigma^{(1)}_{12}$ is given by
\begin{eqnarray}
Tr(W_{1}\sigma^{(1)}_{12})= -0.55
\end{eqnarray}
Since $W_{1}$ is a witness operator and the expectation value is negative for the state $\sigma^{(1)}_{12}$ so the state $\sigma^{(1)}_{12}$ is an entangled state. Now we are in a position to say whether the entangled state is useful for teleportation by calculating its maximum eigenvalue.
The maximum eigenvalue of the state is $\lambda_{max}(\sigma_{12}^{(1)})=0.587>\frac{1}{2}$. Thus the state is useful
in quantum teleportation according to corollary 1. This example is important in the sense that the value of the singlet fraction (after applying SPA-PT operation) is unable to detect the state for possible application of the state $\sigma_{12}^{(1)}$ as a resource state in quantum teleportation but on the other hand, corollary 1 help us to reach the correct conclusion.\\
\subsubsection{Example 2}
Let us consider a quantum state described by the density matrix $\rho_{1}$, which is given by \cite{badziag}
\begin{eqnarray}
\rho_{1}=
\begin{pmatrix}
  0 & 0 & 0 & 0 \\
  0 & \frac{3-2\sqrt{2}}{2} & \frac{1-\sqrt{2}}{2} & 0 \\
  0 & \frac{1-\sqrt{2}}{2} & \frac{1}{2} & 0 \\
  0 & 0 & 0 & \sqrt{2}-1
\end{pmatrix}
\end{eqnarray}
To detect whether the state $\rho_{1}$ is entangled, let us construct a witness operator as\cite{Riccardi}
\begin{eqnarray}
W_{2}=\frac{1}{4} (I \otimes I +\sigma_{z} \otimes I + \sigma_{x} \otimes \sigma_{x}+ \sigma_{y}\otimes \sigma_{y})
\end{eqnarray}
where $\sigma_{i}'s (i=x,y,z)$ are pauli matrices.\\
$Tr(W_{2}\rho_{1})$ can be calculated as
\begin{eqnarray}
Tr(W_{2}\rho_{1})= -1.406
\end{eqnarray}
The negative value of $Tr(W_{2}\rho_{1})$ indicates that the state $\rho_{1}$ is entangled.
Also it can be calculated that the singlet fraction of $\rho_{1}$ is $\frac{1}{2}$. Since $F^{max}(\rho_{1})=\frac{1}{2}$ so it can be concluded that the state $\rho_{1}$ is not useful as a resource state for teleportation. But it is known that all entangled two-qubit mixed states are useful for teleportation. Hence, the inference from singlet fraction that the state $\rho_{1}$ is not useful as a resource state for teleportation is not correct. Let us now apply corollary 1 to detect whether the state $\rho_{1}$ is useful for teleportation. To this end, let us calculate the eigenvalues of $\rho_{1}$ and they are given by \{0.5858,0.4142,0,0\}. The maximum eigenvalue is found out to be $\lambda_{max}(\rho_{1})=0.5858$. Since $\lambda_{max}(\rho_{1})>1/2$, we can conclude that the state $\rho_{1}$ is useful for teleportation.
\subsubsection{Example 3}
Let us take another quantum state from $3\otimes 3$ dimensional Hilbert space described by the density matrix $\rho_{2}$
\begin{eqnarray}
\rho_{2}=
\begin{pmatrix}
  \frac{1-a}{2} & 0 & 0 & 0 & 0 & 0 & 0 & 0 & -0.22 \\
  0 & 0 & 0 & 0 & 0 & 0 & 0 & 0 & 0 \\
  0 & 0 & 0 & 0 & 0 & 0 & 0 & 0 & 0 \\
  0 & 0 & 0 & 0 & 0 & 0 & 0 & 0 & 0 \\
  0 & 0 & 0 & 0 & \frac{1}{2}-a & -0.22 & 0 & 0 & 0 \\
  0 & 0 & 0 & 0 & -0.22 & a & 0 & 0 & 0 \\
  0 & 0 & 0 & 0 & 0 & 0 & 0 & 0 & 0 \\
  0 & 0 & 0 & 0 & 0 & 0 & 0 & 0 & 0 \\
  -0.22 & 0 & 0 & 0 & 0 & 0 & 0 & 0 & \frac{a}{2}
\end{pmatrix}
\end{eqnarray}
where $0.35\leq a \leq 0.369$.\\
The witness operator that detect the state described by the density operator $\rho_{2}$ is given by\cite{Jafarizadeh}
\begin{eqnarray}
W_{3}&=& -\frac{3}{2} A_{3} \otimes A_{3} +\frac{3}{2} S_{3}\otimes S_{3}-\frac{2}{\sqrt{3}}D_{2} \otimes D_{1}+\frac{2}{3}I \otimes D_{1}\nonumber\\&-&\frac{1}{2}D_{1}\otimes D_{1}+\frac{5}{6}D_{2}\otimes D_{2}+\frac{1}{2}iA_{1}\otimes S_{1}+\frac{1}{2}iS_{1}\otimes A_{1}\nonumber\\&-& \frac{1}{2}S_{2} \otimes S_{2}+\frac{1}{2}A_{2} \otimes A_{2}+\frac{2}{3}D_{1}\otimes I- \frac{2}{3} I \otimes I \nonumber\\&+&\frac{2\sqrt{3}}{9}D_{2}\otimes I -\frac{2}{\sqrt{3}}D_{1}\otimes D_{2}+\frac{2\sqrt{3}}{9}I\otimes D_{2}
\end{eqnarray}
where $S_{1}, S_{2}$ and $S_{3}$ are three symmetric Gell-Mann matrices,  $A_{1}, A_{2}$ and $A_{3}$ are three anti-symmetric Gell-Mann matrices and $D_{1}$ and $D_{2}$ are two diagonal Gell-Mann matrices given by
\begin{eqnarray}
S_{1}=
\begin{pmatrix}
0 & 1 & 0  \\
1 & 0 & 0 \\
0 & 0 & 0
\end{pmatrix}, S_{2}=
\begin{pmatrix}
0 & 0 & 1  \\
0 & 0 & 0 \\
1 & 0 & 0
\end{pmatrix}, S_{3}=
\begin{pmatrix}
0 & 0 & 0  \\
0 & 0 & 1 \\
0 & 1 & 0
\end{pmatrix}
\end{eqnarray}
\begin{eqnarray}
A_{1}=
\begin{pmatrix}
0 & -i & 0  \\
i & 0 & 0 \\
0 & 0 & 0
\end{pmatrix}, A_{2}=
\begin{pmatrix}
0 & 0 & -i  \\
0 & 0 & 0 \\
i & 0 & 0
\end{pmatrix}, A_{3}=
\begin{pmatrix}
0 & 0 & 0  \\
0 & 0 & -i \\
0 & i & 0
\end{pmatrix}
\end{eqnarray}
\begin{eqnarray}
D_{1}=
\begin{pmatrix}
1 & 0 & 0  \\
0 & -1 & 0 \\
0 & 0 & 0
\end{pmatrix}, D_{2}=
\begin{pmatrix}
\frac{1}{\sqrt{3}} & 0 & 0  \\
0 & \frac{1}{\sqrt{3}} & 0 \\
0 & 0 & \frac{-2}{\sqrt{3}}
\end{pmatrix}
\end{eqnarray}
We note that $Tr(W_{3}\rho_{2})= 0.44 - 3a$,   where $0.35\leq a \leq 0.369$. Thus, $Tr(W_{3}\rho_{2})<0$ for $0.35\leq a \leq 0.369$.
Hence the state $\rho_{2}$ is an entangled state.\\
Let us calculate the singlet fraction of $\rho_{2}$. To do this, we need maximally entangled basis states in $3\otimes 3$ dimensional
Hilbert space. The maximally entangled basis for two-qutrit system is given by \cite{karimipour}
\begin{eqnarray}
|B_{0}\rangle=\frac{1}{\sqrt{3}}[|00\rangle+|22\rangle-e^{i\frac{\pi}{3}}|11\rangle]\nonumber\\
|B_{1}\rangle=\frac{1}{\sqrt{3}}[|01\rangle+|20\rangle-e^{i\frac{\pi}{3}}|12\rangle]\nonumber\\
|B_{2}\rangle=\frac{1}{\sqrt{3}}[|02\rangle+|21\rangle-e^{i\frac{\pi}{3}}|10\rangle]\nonumber\\
|B_{3}\rangle=\frac{1}{\sqrt{3}}[|11\rangle+|00\rangle-e^{i\frac{\pi}{3}}|22\rangle]\nonumber\\
|B_{4}\rangle=\frac{1}{\sqrt{3}}[|12\rangle+|01\rangle-e^{i\frac{\pi}{3}}|20\rangle]\nonumber\\
|B_{5}\rangle=\frac{1}{\sqrt{3}}[|10\rangle+|02\rangle-e^{i\frac{\pi}{3}}|21\rangle]\nonumber\\
|B_{6}\rangle=\frac{1}{\sqrt{3}}[|11\rangle+|22\rangle-e^{i\frac{\pi}{3}}|00\rangle]\nonumber\\
|B_{7}\rangle=\frac{1}{\sqrt{3}}[|20\rangle+|12\rangle-e^{i\frac{\pi}{3}}|01\rangle]\nonumber\\
|B_{8}\rangle=\frac{1}{\sqrt{3}}[|21\rangle+|10\rangle-e^{i\frac{\pi}{3}}|02\rangle]
\label{mebasis}
\end{eqnarray}
Then the singlet fraction of $\rho_{2}$ can be calculated using the maximally entangled basis (\ref{mebasis}) as
\begin{eqnarray}
F^{max}(\rho_{2})&=& max_{B_{i}}\langle B_{i}|\rho_{2}|B_{i}\rangle,i=0,1,....,8\nonumber\\
&=& \frac{1.22-a}{3}
\label{qutritsingfrac}
\end{eqnarray}
Figure 2 shows that $F^{max}(\rho_{2})$ decreases as the state parameter $a$ increases. The singlet fraction
$F^{max}(\rho_{2})$ is always less than $\frac{1}{3}$ when the state parameter $a$ lying in the interval [0.35,0.369].
Therefore, according to singlet fraction criterion, the state described by the density operator $\rho_{2}$ may or may not be useful in quantum teleportation.\\
\begin{figure}[h!]
\centering
\includegraphics[scale=0.65]{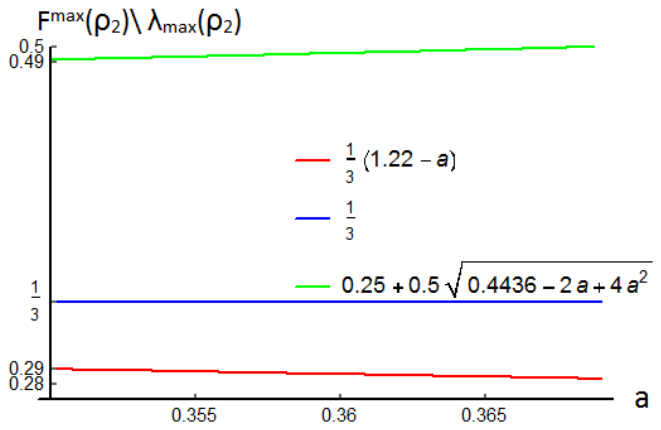}
\caption{Plot of singlet fraction ($F^{max}(\rho_{2})$)/($\lambda_{max}(\rho_{2})$) versus the state parameter a.}
\end{figure}
\noindent Let us now calculate the eigenvalues of $\rho_{2}$. The maximum eigenvalue of $\rho_{2}$ is given by
\begin{eqnarray}
\lambda_{max}(\rho_{2})&=&\frac{1}{4}+\frac{1}{2}\sqrt{0.4436-2a+4a^{2}},\nonumber\\&&
0.35\leq a\leq 0.369
\label{qutritmaxeig}
\end{eqnarray}
We have also shown in Figure 2 that $\lambda_{max}(\rho_{2})$ is always greater than $\frac{1}{3}$ when $a \in [0.35,0.369]$. Thus, our criterion detect that the state
$\rho_{2}$ is useful in quantum teleportation.
\section{Teleportation Criteria in terms of upper bound of the maximum eigenvalue in Dembo's bound}
\noindent In this section, we will study those cases where maximum eigenvalue criterion for teleportation fails i.e. the case where maximum eigenvalue of a given quantum state satisfies the inequality
\begin{eqnarray}
\lambda_{max}(\rho) \leq \frac{1}{d}
\label{ineq2}
\end{eqnarray}
To overcome this problem, we have provided another criterion which is based on Dembo's bound to detect whether the state useful for teleportation or not.\\
Let us consider a qutrit-qutrit system described by the density operator
\begin{eqnarray}
\rho_{3}=
\begin{pmatrix}
\frac{a}{2} & 0 & 0 & 0 & 0 & 0 & 0 & 0 & 0.015\\
0 & \frac{a}{2} & 0 & 0 & 0 & 0 & 0 & 0 & 0 \\
0 & 0 & 0 & 0 & 0 & 0 & 0 & 0 & 0\\
0 & 0 & 0 & 0 & 0 & 0 & 0 & 0 & 0 \\
0 & 0 & 0 & 0 & 0 & 0 & 0 & 0 & 0\\
0 & 0 & 0 & 0 & 0 & 0 & 0 & 0 & 0 \\
0 & 0 & 0 & 0 & 0 & 0 & 0 & 0 & 0 \\
0 & 0 & 0 & 0 & 0 & 0 & 0 & \frac{1-a}{2} & 0 \\
0.015 & 0 & 0 & 0 & 0 & 0 & 0 & 0 & \frac{1-a}{2}
\end{pmatrix}, 0.5 \leq a \leq 0.65
\label{crit1fail}
\end{eqnarray}
We find that $Tr (W_{3}\rho_{3})= -1.53 + 2a <0$, for $0.5\leq a \leq 0.65$. Thus, the witness operator $W_{3}$ detect the state $\rho_{3}$ as an entangled state. But the question is whether the entangled state $\rho_{3}$ is useful in quantum teleportation.\\
Eigenvalues of $\rho_{3}$ are given by: [$0, 0, 0, 0, 0,\frac{1-a}{2},\frac{a}{2}, 0.125(2-\sqrt{16a^{2}-16a+4.0144}), 0.125(2+\sqrt{16a^{2}-16a+4.0144})$]. The maximum eigenvalue is given by $\lambda_{max}(\rho_{3})=0.125(2+\sqrt{16a^{2}-16a+4.0144})$.
We can observe that $\lambda_{max}(\rho_{3})\leq \frac{1}{3}$ When $0.5 \leq a \leq 0.65$. Therefore, our maximum eigenvalue criterion fails to detect whether the state $\rho_{3}$ is useful in teleportation. It motivate us to search for maximal bound of maximum eigenvalue that can be greater than $\frac{1}{d}$ for $d\otimes d$ dimensional system.\\
To start with our search, let us consider the upper bound of maximal eigenvalue of the $d\otimes d$ dimensional quantum state $\rho$ under investigation. The upper bound may be denoted as $\lambda_{max}^{D}(\rho)$ and it is given by R.H.S of the inequality (\ref{db1})
\begin{eqnarray}
\lambda_{max}^{D}(\rho)= \frac{c+\eta_{d^{2}-1}}{2}+\sqrt{\frac{(c-\eta_{d^{2}-1})^{2}}{2}+(b^{*})^{T}b}
\label{qutritmaxeig}
\end{eqnarray}
where  $R_{d^{2}}=\begin{pmatrix}
  R_{d^{2}-1} & b \\
  (b^{*})^{T} & c
  \end{pmatrix}$,
$\eta_{1}$ is the lower bound on minimal eigenvalue of $R_{d^{2}-1}$, $\eta_{d^{2}-1}$ is the upper bound on maximal eigenvalue of $R_{d^{2}-1}$
and $b$ is a vector of dimension $d^{2}-1$.
We are now in a position to provide another criterion in terms of upper bound of maximal eigenvalue of the $d\otimes d$ dimensional quantum state $\rho$.\\
\textbf{Corollary 3:} An arbitrary $d\otimes d$ dimensional mixed NPT entangled state $\rho$ shared between two distant partners is useful as a resource state in quantum teleportation if
\begin{eqnarray}
\lambda_{max}^{D}(\rho)> \frac{1}{d}
\label{critmaxeig}
\end{eqnarray}
\subsubsection{Example 4}
Recalling, the qutrit-qutrit system described by the density operator $\rho_{3}$ given in (\ref{crit1fail}). The maximal bound of maximum eigenvalue of $3\otimes 3$ dimensional density matrix $\rho_{3}$ is given by $\lambda_{max}^{D}(\rho_{3})$. For the $9\times 9$ order density matrix $\rho_{3}$, we have $c = 0.1750$, $(b^{*})^{T}=\begin{pmatrix} 0.015 & 0 & 0 & 0 & 0 & 0 & 0 & 0 \end{pmatrix}$ and $\eta_{8} =0.325$. Using the above values, we find that $\lambda_{max}^{D}(\rho_{3})=0.357$. Thus using corollary 3, we are able to show that the state $\rho_{3}$ is useful in teleportation.
\subsubsection{Example 5}
Let us consider another qutrit-qutrit NPT entangled state, which is given by the density matrix \cite{horodecki3}
\begin{eqnarray}
\rho_{\alpha}= \frac{2}{7}|\phi_{3}^{+}\rangle\langle\phi_{3}^{+}| +\frac{\alpha}{7}\sigma_{+}+\frac{5-\alpha}{7}\sigma_{-}, 4<\alpha \leq 5
\label{dembofail}
\end{eqnarray}
where $|\phi_{3}^{+}\rangle=\frac{1}{\sqrt{3}} \sum_{i=0}^2|ii\rangle$, $\sigma_{+}=\frac{1}{3}(|01\rangle\langle01|+|12\rangle\langle12|+|20\rangle\langle20|)$,
$\sigma_{-}=\frac{1}{3}(|10\rangle\langle10|+|21\rangle\langle21|+|02\rangle\langle02|)$. For the density matrix $\rho_{\alpha}$, we have $c=\frac{2}{21}$, $(b^{*})^{T}$=$\begin{pmatrix} \frac{2}{21} & 0 & 0 & 0 & \frac{2}{21} & 0 & 0 & 0 \end{pmatrix}$ and $\eta_{8}=\frac{5}{21}$. In this case, $\lambda_{max}^{D}(\rho_{\alpha})= 0.3346$, which is greater than $\frac{1}{3}$. Therefore, in this case also we can use corollary 3 to conclude that the state described by the density operator $\rho_{\alpha}$ is useful in quantum teleportation.
\section{Conclusion}
\noindent To summarize, we have modified the relationship between the optimal singlet fraction and partial transpose of a
given state. The modification is required because partial transposition is a non-physical operation and thus it cannot be implemented in the laboratory. We use SPA-PT method to overcome this difficulty and have shown that the modified value of optimal singlet fraction can be estimated in an experiment. Further, we have proposed two criteria based on maximum eigenvalue of the given state and have studied our criteria for the detection of $d\otimes d$ dimensional NPT entangled states useful in quantum teleportation in the given two cases: (i) $F_{max}(\rho)\leq \frac{1}{d}< \lambda_{max}(\rho)$ or (ii) $\lambda_{max}(\rho)\leq \frac{1}{d}< \lambda_{max}^{D}(\rho)$. Our criteria can be realized in an experiment because
maximum eigenvalue can be estimated experimentally \cite{ekert,keyl}.

\section{Acknowledgement}
\noindent A. G. would like to acknowledge the financial support
from CSIR. This work is supported by CSIR File No.
08/133(0035)/2019-EMR-1.

\end{document}